%% file: paper.tex
\documentclass{llncs}

\input{macros}
\input{style}

\declaretheorem[style=theorem]{Fact}

\title{A Predictive Differentially-Private Mechanism for Mobility Traces%
\thanks{This work has been partially supported by the project ANR-12-IS02-001 PACE,
by the INRIA Equipe Associ\'{e}e PRINCESS, by the INRIA Large Scale Initiative
CAPPRIS, and by EU grant agreement no. 295261 (MEALS).}}

\author{%
	Konstantinos Chatzikokolakis\inst{1,2} \and
	Catuscia Palamidessi\inst{2,3} \and
	Marco Stronati\inst{2}
}
\institute{CNRS, France \and LIX, \'{E}cole Polytechnique, France \and INRIA, France}

\allowdisplaybreaks[1]

\begin{document}
\maketitle
\vspace{-10pt}
\begin{abstract}
With the increasing popularity of GPS-enabled handheld devices, location based
applications and services have access to accurate and real-time location
information, raising serious privacy concerns for their millions of users.
Trying to address these issues, the notion of \emph{geo-indistinguishability}
was recently introduced, adapting the well-known concept of Differential Privacy
to the area of location-based systems. A Laplace-based obfuscation mechanism
satisfying this privacy notion works well in the case of a \emph{sporadic} use;
Under repeated use, however, \emph{independently} applying noise leads to a
quick loss of privacy due to the correlation between the location in the trace.

In this paper we show that correlations in the trace can be in fact exploited in
terms of a \emph{prediction function} that tries to guess the new location based
on the previously reported locations. The proposed mechanism tests the quality
of the predicted location using a private test; in case of success the
prediction is reported otherwise the location is sanitized with new noise. If
there is considerable correlation in the input trace, the extra cost of the test
is small compared to the savings in budget, leading to a more efficient
mechanism.

We evaluate the mechanism in the case of a user accessing a location-based
service while moving around in a city. Using a simple prediction function and
two budget spending strategies, optimizing either the utility or the budget
consumption rate, we show that the predictive mechanism can offer substantial
improvements over the independently applied noise.
\end{abstract}

\section{Introduction}
In recent years, the popularity of devices capable of providing an
individual's position with a range of accuracies (e.g.
wifi-hotspots, GPS, etc) has led to a growing use of ``location-based systems''
that record and process location data. A typical example of such systems are
Location Based Services (LBSs) -- such as mapping applications,
Points of Interest retrieval, coupon providers, GPS navigation, and
location-aware social networks -- providing a service
related to the user's location.
Although users are often willing to disclose their location in order to obtain a
service, there are serious concerns about the privacy implications of the
constant disclosure of location information. 

In this paper we consider the problem of a user accessing a LBS while wishing
to hide his location from the service provider. We should emphasize that, in
contrast to several works in the literature
\cite{Gruteser:03:MobiSys,Mokbel:06:VLDB}, we are interested not in hiding the
user's \emph{identity}, but instead his \emph{location}. In fact, the user might
be actually authenticated to the provider, in order to obtain a personalized
service (personalized recommendations, friend information from a social network,
etc); still he wishes to keep his location hidden.

Several techniques to address this problem have been proposed in the literature,
satisfying a variety of location privacy definitions. A widely-used such notion
is $k$-anonymity (often called $l$-diversity in this context), requiring that
the user's location is indistinguishable among a set of $k$ points. This could
be achieved either by adding \emph{dummy locations} to the query
\cite{Kido:05:ICDE,Shankar:09:UbiComp}, or by creating a \emph{cloaking region}
including $k$ locations with some semantic property, and querying the service
provider for that cloaking region
\cite{Bamba:08:WWW,Duckham:05:Pervasive,Xue:09:LoCa}. 
A different approach is to report an \emph{obfuscated} location $z$ to the service
provider, typically obtained by adding random noise to the real one. Shokri et
al. \cite{Shokri:12:CCS} propose a method to construct an obfuscation mechanism
of optimal privacy for a given quality loss constraint, where privacy is
measured as the expected error of a Bayesian adversary trying to guess the
user's location \cite{Shokri:11:SP}.

The main drawback of the aforementioned location privacy definitions is that they
depend on the adversary's background knowledge, typically modeled as a prior
distribution on the set of possible locations. If the adversary can rule out
some locations based on his prior knowledge, then $k$-anonymity will be
trivially violated. Similarly, the adversary's expected error directly depends
on his prior. As a consequence, these definitions give no precise guarantees
in the case when the adversary's prior is different.

Differential privacy \cite{Dwork:06:ICALP} was introduced for statistical
databases exactly to cope with the issue of prior knowledge. The goal in this
context is to answer aggregate queries about a group of individuals without
disclosing any individual's value. This is achieved by adding random noise to
the query, and requiring that, when executed on two databases $x,x'$ differing
on a single individual, a mechanism should produce the same answer $z$ with
similar probabilities. Differential privacy has been successfully used in the
context of location-based systems
\cite{Machanavajjhala:08:ICDE,Ho:11:GIS,Chen:12:CCS} when \emph{aggregate}
location information about a large number of individuals is published. 
However,
in the case of a single individual accessing an LBS, this property is too
strong, as it would require the information sent to the provider to be
independent from the user's location.

Our work is based on ``\emph{geo-indistinguishability}'', a variant of
differential privacy adapted to location-based systems, introduced recently in
\cite{Andres:13:CCS}. Based on the idea that the user should enjoy strong
privacy within a small radius, and weaker as we move away from his real
location, geo-indistinguishability requires that the closer (geographically) two
locations are, the more indistinguishable they should be. This means that when
locations $x,x'$ are close they should produce the same reported location $z$
with similar probabilities; however the probabilities can become substantially
different as the distance between $x$ and $x'$ increases. This property can be
achieved by adding noise to the user's location drawn from a 2-dimensional
Laplace distribution.

In practice, however, a user rarely performs a \emph{single} location-based
query. As a motivating example, we consider a user in a city performing
different activities throughout the day: for instance he might have lunch, do
some shopping, visit friends, etc. During these activities, the user performs
several queries: searching for restaurants, getting driving directions, finding
friends nearby, and so on. For each query, a new obfuscated location needs to be
reported to the service provider, which can be easily obtained by independently
adding noise at the moment when each query is executed. We refer to
independently applying noise to each location as the
\emph{independent mechanism}.

However, it is easy to see that privacy is degraded as the number of queries
increases, due to the \emph{correlation} between the locations. Intuitively, in
the extreme case when the user never moves (i.e. there is perfect
correlation), the reported locations are centered around the real one,
completely revealing it as the number of queries increases. Technically,
the independent mechanism applying $\epsilon$-geo-indistinguishable noise
(where $\epsilon$ is a privacy parameter) to $n$ location can be shown to
satisfy $n\epsilon$-geo-indistinguishability \cite{Andres:13:CCS}. This is
typical in the area of differential privacy, in which $\epsilon$ is
thought as a privacy \emph{budget}, consumed by each query; this linear
increase makes the mechanism applicable only when the number of queries remains
small. Note that any obfuscation mechanism is bound to cause privacy loss when
used repeatedly; geo-indistinguishability has the advantage of directly
quantifying this loss terms of the consumed budget.

The goal of this paper is to develop a \emph{trace obfuscation} mechanism with a
smaller \emph{budget consumption rate} than applying independent noise. The main
idea is to actually use the correlation between locations in the trace to our
advantage. Due to this correlation, we can often \emph{predict} a point close to
the user's actual location from information previously revealed. 
For instance, when the user performs multiple different queries from
the same location - e.g. first asking for shops and later for
restaurants - we could intuitively use the same reported location in
all of them, instead of generating a new one each time.
However, this implicitly reveals that the user is not
moving, which violates geo-indistinguishability (nearby locations produce
completely different observations); hence the decision to report the same
location needs to be done in a private way.

Our main contribution is a \emph{predictive mechanism} with three
components: a \emph{prediction function} $\Omega$, a \emph{noise mechanism} $N$ and a
\emph{test mechanism} $\Theta$. The mechanism behaves as follows: first, the list of
previously reported locations (i.e. information which is already public) are
given to the prediction function, which outputs a predicted location $\predz$.
Then, it tests whether $\predz$ is within some threshold $l$ from the user's
current location using the test mechanism. The test itself should be private:
nearby locations should pass the test with similar probabilities. If the test
succeeds then $\predz$ is reported, otherwise a new reported location is
generated using the noise mechanism.

The advantage of the predictive mechanism is that the budget is consumed only
when the test or noise mechanisms are used. Hence, if the prediction rate is
high, then we will only need to pay for the test, which can be substantially
cheaper in terms of budget. The configuration of $N$ and $\Theta$ is done via a
\emph{budget manager} which decides at each step how much budget to spend on
each mechanism. The budget manager is also allowed to completely skip the test
and blindly accept or reject the prediction, thus saving the corresponding
budget. The flexibility of the budget manager allows for a dynamic behavior,
constantly adapted to the mechanism's previous performance. We examine in detail
two possible budget manager strategies, one maximizing utility under a fixed
budget consumption rate and one doing the exact opposite, and explain in detail
how they can be configured.

Note that, although we exploit correlation for efficiency, the predictive
mechanism is shown to be private independently from the prior distribution on
the set of traces. If the prior presents correlation, and the
prediction function takes advantage of it, the mechanism can
achieve a good budget consumption rate, which translates either to better
utility or to a greater number of reported points than the independent
mechanism. If there is no correlation, or the prediction does not take advantage
of it, then the budget consumption can be worse than the independent mechanism.
Still, thanks to the arbitrary choice of the prediction function and the budget
manager, the predictive mechanism is a powerful tool that can be adapted to a
variety of practical scenarios.

We experimentally verify the effectiveness of the mechanism on our
motivating example of a user performing various activities in a city,
using two large data sets of GPS trajectories in the Beijing urban area
(\cite{DBLP:journals/debu/ZhengXM10,DBLP:conf/gis/YuanZZXXSH10}).
Geolife \cite{DBLP:journals/debu/ZhengXM10} collects the movements of
several users, using a variety of transportation means, including
walking, while in Tdrive \cite{DBLP:conf/gis/YuanZZXXSH10} we find
exclusively taxi drivers trajectories.
The results for both budget managers, with and without the skip strategy, show
considerable improvements with respect to independently applied noise. More
specifically, we are able to decrease average error up to 40\% and budget
consumption rate up to 64\%. The improvements are significative enough to
broaden the applicability of geo-indistinguishability to cases impossible
before: in our experiments we cover 30 queries with reasonable error which is
enough for a full day of usage; alternatively we can drive the error down from 5
km to 3 km, which make it acceptable for a variety of application.

Note that our mechanism can be efficiently implemented on the user's phone,
and does not require any modification on the side of the provider, hence it can
be seamlessly integrated with existing LBSs.

\vspace{-5pt}
\paragraph{Contributions}
The paper's contributions are the following:
\begin{itemize}[noitemsep,topsep=0pt,parsep=0pt,partopsep=0pt]
\item We propose a predictive mechanism that exploits
	correlations on the input by means of a prediction function.
\item We show that the proposed mechanism is private and
  provide a bound on its utility.
\item We instantiate the predictive mechanism for
  location privacy, defining a prediction function and two
  budget managers, optimizing utility and budget consumption rate.
\item We evaluate the mechanism on two large sets of
  GPS trajectories and confirm our design goals, showing substantial
  improvements compared to independent noise.
\end{itemize}

A final note on the generality of out method, the work presented in
this paper started with the objective to extend the use of
geo-indistinguishability to location traces with a more efficient use
of the budget but thanks to the generality of the approach it
developed into a viable mechanism for other domains in the same family
of metric based \priv{\dx}.
It is indeed the major focus of our future work to apply this
technique in new fields such as smart meters and back in the standard
domain of differential privacy, statistical databases.

\vspace{-5pt}
\paragraph{Plan of the paper}
In the next section we recall some preliminary notions about
differential privacy  and geo-indistinguishability.
In Section \ref{sec:predictive-mechanism} we present in detail the
components of the predictive mechanism, including budget managers and
skip strategies, together with the main results of privacy and
utility.
In Section \ref{sec:predictive-geo-ind} we apply the predictive
mechanism to location privacy, defining a prediction
function, skip strategies and detailed configurations of the budget manager.
Finally in Section \ref{sec:case-study} we describe the experiments
and their results. All proofs can be found in the appendix.

\vspace{-5pt}
\section{Preliminaries}\label{sec:preliminaries}

We briefly recall here some useful notions from the literature.

\vspace{-5pt}
\subsubsection{Differential Privacy and Geo-indistinguishability.}

The privacy definitions used in this paper are based on a generalized variant of
differential privacy that can be defined on an arbitrary set of secrets $\calx$
(not necessarily on databases), equipped with a metric $\dx$
\cite{Reed:10:ICFP,Chatzikokolakis:13:PETS}. The distance $\dx(x,x')$ expresses
the \emph{distinguishability level} between the secrets $x$ and $x'$, modeling
the privacy notion that we want to achieve. A small value denotes that the
secrets should remain indistinguishable, while a large value means that we allow
the adversary to distinguish them.

Let $\calz$ be a set of \emph{reported values} and let $\calp(\calz)$ denote the
set of probability measures over $\calz$. The multiplicative distance $\dprob$ on
$\calp(\calz)$ is defined as
$d_{\cal P}(\mu_1,\mu_2) = \sup_{Z \subseteq \calz} |\ln \frac{\mu_1(Z)}{\mu_2(Z)}|$
with $|\ln \frac{\mu_1(Z)}{\mu_2(Z)}|=0$ if
both $\mu_1(Z),\mu_2(Z)$ are zero and $\infty$ if only one of them is zero.
Intuitively $\dprob(\mu_1,\mu_2)$ is small if $\mu_1,\mu_2$ assign similar probabilities
to each reported value.

A mechanism is a (probabilistic) function $K:\calx\to\calp(\calz)$, assigning to
each secret $x$ a probability distribution $K(x)$ over the reported values.
The generalized variant of differential privacy, called \priv{\dx}, is defined
as follows:
\vspace{-3pt}
\begin{definition}[\priv{\dx}]
  A mechanism $K : \calx\rightarrow \calp(\calz)$ satisfies
  \priv{\dx} iff:
  \[
	  \dprob(K(x),K(x')) \leq \dx(x,x')
		\qquad \forall x,x' \in \calx
  \]
  or equivalently
  $ K(x)(Z) \leq e^{\dx (x,x')} K(x')(Z)
  \quad \forall x,x'\in\calx,Z \subseteq \calz
  $.
\end{definition}
Different choices of $\dx$ give rise to different privacy notions; it is also
common to scale our metric of interest by a privacy parameter $\epsilon$ (note
that $\epsilon\dx$ is itself a metric).

The most well-known case is when $\calx$ is a set of databases with the hamming
metric $\hamming(x,x')$, defined as the number of rows in which $x,x'$ differ.
In this case \priv{\epsilon\hamming} is the same as \edpold{}, requiring that
for adjacent $x,x'$ (i.e. differing on a single row) $\dprob(K(x),K(x')) \leq
\epsilon$. Moreover, various other privacy notions of interest can be captured
by different metrics \cite{Chatzikokolakis:13:PETS}.

\vspace{-5pt}
\paragraph{Geo-indistinguishability}
In the case of location privacy, which is the main motivation of this paper, the
secrets $\calx$ as well as the reported values $\calz$ are sets of locations
(i.e. subsets of $\reals^2$), while $K$ is an obfuscation mechanism. Using the
Euclidean metric $\euclid$, we obtain \priv{\epsilon\euclid}, a natural notion
of location privacy called \geoind{} in \cite{Andres:13:CCS}. This privacy
definition requires that the closer (geographically) two location are, the more
similar the probability of producing the same reported location $z$ should be.
As a consequence, the service provider is not allowed to infer the user's
location with accuracy, but he can get approximate information required to
provide the service.

Seeing it from a slightly different viewpoint, this notion offers privacy
\emph{within any radius $r$} from the user, with a level of distinguishability
$\epsilon r$, proportional to $r$. Hence, within a small radius the user enjoys
strong privacy, while his privacy decreases as $r$ gets larger. This gives us
the flexibility to adjust the definition to a particular application: typically
we start with a radius $r^*$ for which we want strong privacy, which can range
from a few meters to several kilometers (of course a larger radius will lead to
more noise). For this radius we pick a relatively small $\epsilon^*$ (for
instance in the range from $\ln 2$ to $\ln 10$), and set $\epsilon = \epsilon^*
/ r^*$. Moreover, we are also flexible in selecting a different metric between
locations, for instance the Manhattan or a map-based distance.

Two characterization results are also given in \cite{Andres:13:CCS},
providing intuitive interpretations of geo-indistinguishability. Finally, it
is shown that this notion can be achieved by adding noise from a 2-dimensional
Laplace distribution.

\vspace{-5pt}
\subsubsection*{Protecting location traces.}

\floatstyle{plain}
\restylefloat{figure}
\begin{wrapfigure}[10]{L}{0.32\textwidth}
\vspace{-20pt}

\begin{lstlisting}[mathescape, language=C, emph={mechanism},emphstyle=\textbf]
mechanism $\IM$($\xb$)
  $\zb := \emptytrace$
  for $i := 1$ to $|\xb|$
    $z := N(\epsilon_N)(\xb[i])$
    $\zb := z :: \zb$
  return $\zb$
\end{lstlisting}
\caption{Independent Mechanism}\label{fig:IM}
\end{wrapfigure}

Having established a privacy notion for single locations, it is natural to
extend it to location \emph{traces} (sometimes called \emph{trajectories} in the
literature). Although location privacy is our main interest,
this can be done for traces having any secrets with a corresponding metric as
elements. We denote by $\xb =[x_1,\ldots,x_n]$ a trace, by $\xb[i]$ the $i$-th
element of $\xb$, by $\emptytrace$ the empty trace and by $x :: \xb$ the trace
obtained by adding $x$ to the head of $\xb$. 
We also define $\mathtt{tail}(x :: \xb) = \xb$. 
To obtain a privacy notion, we need to define an appropriate
metric between traces. A natural choice is the maximum metric
$\dmax(\vc{x},\vc{x}') = \max_i \dx(\xb[i],\xb'[i])$.
This captures the idea that two traces are as distinguishable as their most
distinguishable points. In terms of protection within a radius, if $\xb$ is
within a radius $r$ from $\xb'$ it means that $\xb[i]$ is within a radius $r$
from $\xb'[i]$. Hence, \priv{\epsilon\dmax} ensures that all secrets are
protected within a radius $r$ with the same distinguishability level $\epsilon
r$.

In order to sanitize $\xb$ we can simply apply a \emph{noise mechanism}
independently to each secret $x_i$. We assume that a family of noise mechanisms
$N(\epsilon_N): \calx\to\calp(\calz)$ are available, parametrized by $\epsilon_N>0$,
where each mechanism $N(\epsilon_N)$ satisfies \priv{\epsilon_N}. The resulting
mechanism, called the \emph{independent mechanism} $\IM:
\calx^n\to\calp(\calz^n)$, is shown in Figure~\ref{fig:IM}. As explained in the
introduction, the main issue with this approach is that
$\IM$ is \privadj{n\epsilon\dmax}, that is, the budget consumed increases
linearly with $n$.

\vspace{-5pt}
\subsubsection*{Utility.}

The goal of a privacy mechanism is not to hide completely the secret but to
disclose enough information to be useful for some service while hiding the rest
to protect the user's privacy. Typically these two requirements go in opposite
directions: a stronger privacy level requires more noise which results in a
lower utility.

Utility is a notion very dependent on the application we target; to measure
utility we start by defining a notion of \emph{error}, that is a distance
$\derr$ between a trace $\xb$ and a sanitized trace $\zb$.
In the case of location-based systems we want to report locations as close as
possible to the original ones, so a natural choice is to define the error
as the average geographical distance between the locations in the trace:
\begin{equation}
	\derr(\vc{x},\vc{z}) = \smallfrac{1}{|\vc x|}\smallsum{i} \euclid(\vc x [i],\vc z [i])\label{eq:4}
\end{equation}

We can then measure the utility of a trace obfuscation mechanism $K:\calx^n\to\calp(\calz^n)$
by the \emph{average-case} error, defined as the expected value of $\derr$:
\[
	E[\derr] = \displaystyle
		\smallsum{\vc{x}} \pi(\vc{x}) 
		\smallsum{\vc{z}} K(\vc{x})(\vc{z}) \; \derr(\vc{x},\vc{z})
\]
where $\pi \in\calp(\calx^n)$ is a prior distribution on traces.

On the other hand, the worst-case error is usually unbounded,
since typical noise mechanisms (for instance the Laplace one)
can return values at arbitrary distance from the original one.
Hence, we are usually interested in the $p$-th percentile
of the error, commonly expressed in the form of $\alpha(\delta)$-\emph{accuracy}
\cite{Roth:10:STOC}.
A mechanism $K$ is $\alpha(\delta)$-accurate iff for all $\delta$:
$Pr[ \derr(\vc{x},\vc{z}) \le \alpha(\delta) ] \ge \delta$.
In the rest of the paper we will refer to $\alpha(0.9)$ (or simply $\alpha$) as the
``worst-case'' error.

Note that in general, both $E[\derr]$ and $\alpha(\delta)$ depend
on the prior distribution $\pi$ on traces. However, due to the mechanism's
symmetry, the utility of the Laplace mechanism is independent from the prior,
and as a result, the utility of the independent mechanism (using the Laplace as
the underlying noise mechanism) is also prior-independent. On the other
hand, the utility of the predictive mechanism, described in the next section,
will be highly dependent on the prior. As explained in the introduction, the
mechanism takes advantage of the correlation between the points in the trace
(a property of the prior), the higher the correlation the better
utility it will provide.

\vspace{-5pt}
\section{A predictive $\dx$-private mechanism}\label{sec:predictive-mechanism}

We are now ready to introduce our prediction-based mechanism. Although our main
motivation is location privacy, the mechanism can work for traces of any secrets
$\calx$, equipped with a metric $\dx$. The fundamental intuition of our work is
that the presence of correlation on the secret can be exploited to the advantage
of the mechanism. A simple way of doing this is to try to predict new secrets
from past information; if the secret can be predicted with enough accuracy it is
called \emph{easy}; in this case the prediction can be reported without adding
new noise. One the other hand, \emph{hard} secrets, that is those that cannot be
predicted, are sanitized with new noise. Note the difference with the
independent mechanism where each secret is treated independently from the
others.

Let $\calb = \{0,1\}$. A boolean $b \in \calb$ denotes whether a point is easy
(0) or hard (1). A sequence $\rb = [z_1,b_1,\ldots,z_n,b_n]$ of reported values and
booleans is called a \emph{run}; the set of all runs is denoted by $\calr =
(\set{Z} \times \set{B})^*$. A run will be the output of our predictive
mechanism; note that the booleans $b_i$ are considered public and will be
reported by the mechanism.

\vspace{-5pt}
\paragraph{Main components}
The predictive mechanism has three main components: first, the \emph{prediction}
is a deterministic function $\Omega: \set{R} \rightarrow \set{Z}$, taking as
input the run reported up to this moment and trying to predict the next
\emph{reported value}. The output of the prediction function is denoted by
${\tilde z} = \Omega(\vc{r})$. Note that, although it is natural to think of
$\Omega$ as trying to predict the secret, in fact what we are trying to predict is
the reported value. In the case of location privacy, for instance, we want to
predict a reported location at acceptable distance from the actual one.
Thus, the possibility of a successful prediction should not be viewed as a
privacy violation.

Second, a \emph{test} is a family of mechanisms
$\Theta(\epsilon_{\theta},l,\tilde{z}): \set{X} \rightarrow \set{P}(\set{B})$,
parametrized by $\epsilon_\theta,l,\tilde{z}$. The test takes as input the
secret $x$ and reports whether the prediction $\tilde{z}$ is acceptable or not
for this secret. If the test is successful then the prediction will be used
instead of generating new noise. The purpose of the test is to guarantee a
certain level of utility: predictions that are farther than the threshold $l$
should be rejected. Since the test is accessing the secret, it should be private
itself, where $\epsilon_\theta$ is the budget that is allowed to be spent for
testing.

The test mechanism that will be used throughout the paper is the one below,
which is based on adding Laplace noise to the threshold $l$:
\vspace{-5pt}
\begin{equation}
	\Theta (\epsilon_{\theta},l,\tilde{z})(x) = \left\{
	  \begin{array}{ll}
		0 & \text{if } \dx(x,\tilde z) \leq l + Lap(\epsilon_{\theta})\\
		1 & \text{ow.}
	  \end{array}
	\right.
	\label{eq:laplacetest}
\end{equation}
\vspace{-8pt}

The test is defined for all $\epsilon_\theta > 0, l \in [0,+\infty),
\tilde{z}\in\calz$,
and can be used for any metric $\dx$, as long as the domain of reported
values is the same as the one of the secrets (which is the case for location
obfuscation) so that $\dx(x,\tilde z)$ is well defined.

Finally, a \emph{noise mechanism} is a family of mechanisms $N(\epsilon_N):
\calx\to\calp(\calz)$, parametrized by the available budget $\epsilon_N$.
The noise mechanism is used for hard secrets that
cannot be predicted.

\vspace{-5pt}
\paragraph{Budget management}
The parameters of the mechanism's components need to be configured at each step.
This can be done in a dynamic way using the concept of a
\emph{budget manager}.
A budget manager $\beta$ is a function that takes as input the run
produced so far and returns the budget and the threshold to be used
for the test at this step as well as the budget for the noise
mechanism: $\beta(\vc{r}) = (\epsilon_{\theta},\epsilon_N,l)$.
We will also use $\beta_{\theta}$ and $\beta_N$ as shorthands to get
just the first or the second element of the result.

Of course the amount of budget used for the test should always be less than the
amount devoted to the noise, otherwise it would be more convenient to just use
the independent noise mechanism. Still, there is great flexibility in
configuring the various parameters and several strategies can be implemented
in terms of a budget manager. 
In this work we fix the level of privacy guaranteed, as it is our
priority, and for predictable traces the budget manager will improve
the utility, in terms of average error or budget consumption rate.

In the next section we will discuss two possible
budget management policies, one maximizing utility under a fixed
budget consumption rate and one doing the exact opposite.

All the components are defined here with the minimal information
needed for their function, consider though that all of them could
access additional public information, for example we may want to
enrich the prediction function for a database with common statistics
of a population or in geolocalization with maps of the territory.

\vspace{-5pt}
\paragraph{The mechanism}

\setlength{\belowcaptionskip}{-12pt}
\begin{figure*}[t]
\centering
\begin{subfigure}[b]{0.45\textwidth}


\begin{lstlisting}[mathescape, language=C, emph={mechanism,to},emphstyle=\textbf]
mechanism $\PM$($\xb$)
  $\rb := \emptytrace$
  for $i := 1$ to $|\xb|$
    $(z,b) := \Step(\rb)(\xb[i])$
    $\rb := (z,b) :: \rb$
  return $\rb$
\end{lstlisting}
\caption{Predictive Mechanism}\label{fig:PM}
\end{subfigure}
~
\begin{subfigure}[b]{0.45\textwidth}

\begin{lstlisting}[mathescape, language=C, emph={mechanism,then},emphstyle=\textbf]
mechanism $\Step(\rb)$($x$)
  $(\epsilon_{\theta},\epsilon_N,l) := \beta(\vc{r})$
  $\tilde{z} := \Omega(\vc{r})$
  $b := \Theta(\epsilon_{\theta},l,\tilde{z})(x)$
  if $b == 0$ then $z := \tilde{z}$
  else $z := N(\epsilon_N)(x)$
  return $(z,b)$
\end{lstlisting}
\caption{Single step of the Predictive Mechanism}\label{fig:K}
\end{subfigure}
\end{figure*}
\setlength{\belowcaptionskip}{0pt}

We are now ready to fully describe our mechanism.
A single step of the predictive mechanism, displayed in Figure~\ref{fig:K}, is
a family of mechanisms $\Step(\vc{r}): \set{X} \rightarrow \set{P}(\set{Z}
\times \set{B})$, parametrized by the run $\rb$ reported up to this point.
The mechanism takes a secret $x$ and returns a reported value $z$, as well as a
boolean $b$ denoting whether the secret was easy or hard.
First, the mechanism obtains the various configuration parameters from the
budget manager as well as a prediction $\tilde{z}$. Then the prediction
is tested using the test mechanism. If the test is successful the prediction
is returned, otherwise a new reported value is generated using the noise
mechanism. 

Finally, the predictive mechanism, displayed in Figure~\ref{fig:PM}, is a
mechanism $\PM: \set{X}^n \rightarrow \set{P}(\set{R})$. It takes as input a
trace $\xb$, and applies $\Step(\rb)$ to each secret, while extending at each
step the run $\rb$ with the new reported values $(z,b)$.

Note that an important advantage of the mechanism is that it is \emph{online},
that is the sanitization of each secret does not depend on future secrets. This
means that the user can query at any time during the life of the system, as
opposed to \emph{offline} mechanisms were all the queries need to be asked
before the sanitization.
Furthermore the mechanism is \emph{dynamic}, in the sense that the
secret can change over time (e.g. the position of the user) contrary to
static mechanism where the secret is fixed (e.g. a static database).

It should be also noted that, when the user runs out of budget, he should
in principle stop using the system. This is typical in the area of
differential privacy where a database should not being queried after the
budget is exhausted. In practice, of course, this is not realistic, and new
queries can be allowed by resetting the budget, essentially assuming either
that there is no correlation between the old and new data, or that the
correlation is weak and cannot be exploited by the adversary. In the case
of location privacy we could, for instance, reset the budget at the end of
each day. We are currently investigating proper
assumptions under which the budget can be reset while satisfying a formal
privacy guarantee. The question of resetting the budget is
open in the field of differential privacy and
is orthogonal to our goal of making an
efficient use of it.

The main innovation of this mechanism if the use of the prediction
function, which allows to decouple the privacy mechanism from the
correlation analysis, creating a family of modular mechanisms where by
\emph{plugging} in different predictions (or updating the existing) we
are able to work in new domains.
Moreover proving desirable security properties about the mechanism
independently of the complex engineering aspects of the prediction is
both easier and more reliable, as shown in the next sections.

\vspace{-8pt}
\subsection{Privacy}
We now proceed to show that the predictive mechanism described in the
previous section is \privadj{\dx}.
The privacy of the predictive mechanism depends on that of its components.
In the following, we assume that each member of the families of test and noise
mechanisms is \privadj{\dx} for the corresponding privacy parameter:
\begin{equation}
    \forall \epsilon_{\theta},l,\tilde{z}. \;
	\Theta(\epsilon_{\theta},l,\tilde{z})
	\; \text{is $\epsilon_{\theta} \dx$-private} \label{privacy-assumption-test}
\end{equation}
\vspace{-15pt}
\begin{equation}
    \;\forall \epsilon_N. \quad \;\; 
	N(\epsilon_N) 
	\quad \;\;\text{is $\epsilon_N \dx$-private} \label{privacy-assumption-noise}
\end{equation}
In the case of the test $\Theta(\epsilon_{\theta},l,\tilde{z})$ defined in
\eqref{eq:laplacetest}, we can
show that it is indeed \privadj{\dx}, independently of the metric or threshold
used.

\begin{Fact}[Privacy of Test function]
	The family of test mechanisms $\Theta(\epsilon_{\theta},l,\tilde{z})$
	defined by \eqref{eq:laplacetest} satisfies
 	assumption \ref{privacy-assumption-test}.
\end{Fact}

The global budget for a certain run $\vc{r}$ using a budget manager $\beta$
is defined as:
\begin{equation}
	\epsilon_{\beta}(\vc{r}) = \left\{
	  \begin{array}{lll}
		0 && \text{if } |\vc r|=0 \\
		\beta_{\theta}(\vc{r}) + b(\vc r) \times \beta_N(\vc{r}) 
		+\; \epsilon_{\beta}(\mathtt{tail}(\vc r)) && o.w.
	  \end{array}
	\right.\label{epsilon-run}
\end{equation}
As already discussed, a hard step is more expensive than an easy
step because of the cost of the noise mechanism.

Building on the privacy properties of its components, we first show that
the predictive mechanism satisfies a property similar to \priv{\dx}, with
a parameter $\epsilon$ that depends on the run.

\begin{lemma}\label{th:privacy}
	Under the assumptions
	\eqref{privacy-assumption-test},\eqref{privacy-assumption-noise},
	for the test and noise mechanisms,
  the predictive mechanism $\PM$, using the
  budget manager $\beta$, satisfies
  \begin{equation}
    \PM(\vc{x})(\vc{r}) \leq 
    e^{\epsilon_{\beta}(\vc{r}) \; \dmax(\vc{x}, \vc{x'})} \PM(\vc{x'})(\vc{r})
	\qquad \forall \vc{r},\vc{x},\vc{x'}
  \end{equation}
\end{lemma}
This results shows that there is a difference between the budget spent on a
``good'' run, where the input has a considerable correlation, the prediction
performs well and the majority of steps are easy, and a run with uncorrelated
secrets, where any prediction is useless and all the steps are hard. In the
latter case it is clear that our mechanism wastes part of its budget on tests
that always fail, performing worse than an independent mechanism.

Finally, the overall privacy of the mechanism will depend on the budget
spent on the worst possible run.

\begin{theorem}[\priv{\dx}] 
	Under the assumptions
	\eqref{privacy-assumption-test},\eqref{privacy-assumption-noise},
	for the test and noise mechanisms,
	the predictive mechanism $\PM$, using the
	budget manager $\beta$, satisfies \priv{\epsilon\dmax}, with \;
	$\epsilon = \sup_{\vc{r}} \; \epsilon_{\beta}(\vc{r})$.
\end{theorem}

Based on the above result, we will use \emph{$\epsilon$-bounded} budget
managers, imposing an overall budget limit $\epsilon$ independently from the
run. Such a budget manager provides a fixed privacy guarantee by sacrificing
utility: in the case of a bad run it either needs to lower the budget spend per
secret, leading to more noise, or to stop early, handling a smaller number of
queries. In practice, however, using a
prediction function tailored to a specific type of correlation we can achieve good
efficiency. 
Moreover, we have the flexibility to use several prediction
functions, each specialized on a specific set of correlated inputs, and to
dynamically switch off the prediction in case it performs poorly (see
Section~\ref{sec:skipping-test}).

\vspace{-8pt}
\subsection{Utility}

We now turn our attention to the utility provided by the predictive mechanism.
The property we want to prove is $\alpha(\delta)$-\emph{accuracy},
introduced in Section~\ref{sec:preliminaries}. Similarly to the case of
privacy, the accuracy of the predictive mechanism depends on that of its
components, that is, on the accuracy of the noise mechanism, as well as
the one of the Laplace mechanism employed by the test
$\Theta(\epsilon_\theta,l,\tilde{z})$ 
\eqref{eq:laplacetest}. We can now state a result about the utility of a
\emph{single step} of the predictive mechanism.
\vspace{-5pt}
\begin{proposition}[accuracy]\label{th:accuracy}
  Let $\vc{r}$ be a run, $\beta$ a budget manager, let
  $(\epsilon_{\theta}, \epsilon_N, l) = \beta(\vc{r})$ and let
  $\alpha_N(\delta)$, $\alpha_{\theta}(\delta)$ be the accuracy of
  $N(\epsilon_N)$, $Lap(\epsilon_\theta)$ respectively.
  Then the accuracy of $\Step(\vc{r})$ is
  $\alpha(\delta) = \max(\alpha_N(\delta), l + \alpha_{\theta}(\delta))$
\end{proposition}
\vspace{-3pt}

This result provides a bound for the accuracy of the predictive mechanism at
each step. The bound depends on the triplet used
$(\epsilon_{\theta},\epsilon_N,l)$ to configure the test and noise mechanisms
which may vary at each step depending on the budget manager used, thus the bound
is step-wise and may change during the use of the system. 

It should be noted that the bound is independent from the prediction
function used, and assumes that the prediction gives the worst possible accuracy
allowed by the test. Hence, under a prediction that always fails the bound is
tight; however, under an accurate prediction function, the  mechanism can achieve
much better utility, as shown in the evaluation of Section \ref{sec:case-study}.

As a consequence, when we configure the
mechanism in Section~\ref{sec:configuration}), we scale down this bound to
account for the improvement due to the prediction.

In the next section we will discuss the possibility to skip entirely
the test in certain cases, of course our bound on accuracy cannot hold
is such a case unless the mechanism designer can provide some safe
assumptions on the accuracy of its skip-the-test strategy.

\vspace{-8pt}
\subsection{Skipping the test}\label{sec:skipping-test}
The amount of budget devoted to the test is still linear in the number
of steps and can amount to a considerable fraction; for this reason,
given some particular conditions, we may want to skip it altogether
using directly the prediction or the noise mechanism.
The test mechanism we use \eqref{eq:laplacetest} is defined
for all
$\epsilon_\theta > 0, l \in [0,+\infty)$.
We can extend it to the case
$\epsilon_\theta = 0, l \in \{-\infty,+\infty \}$
with the convention
that $\Theta(0,+\infty,\tilde{z})$ always returns 1 and
$\Theta(0,-\infty,\tilde{z})$ always returns 0. 
This convention is based on the
intuition that $\dx(x,\tilde{z})$ is always greater than $-\infty$ and smaller
than $+\infty$, and no budget is needed to test this.

The new test mechanisms are independent of the input $x$ so they can be
trivially shown to be private, with no budget being consumed.
\begin{Fact}[Privacy of Test function]
  The test functions $\Theta(0,+\infty,\tilde{z})$ and \\
  $\Theta(0,-\infty,\tilde{z})$
  satisfy assumption \ref{privacy-assumption-test}.
\end{Fact}

Now if $\beta$ returns $(0, \epsilon_N, -\infty)$ we always fallback to the
noise mechanism $N(\epsilon_N)$; this is especially useful when we know the
prediction is not in conditions to perform well and testing would be a waste of
budget. For instance, consider a prediction function that needs at least a
certain number $n$ of previous observables to be able to predict with enough
accuracy; in this case we can save some budget if we directly use the noise
mechanism for those $n$ steps without testing.
Note that the bound on utility is preserved in this case, as we can rely on
the $\alpha_N(\delta)$-accuracy of $N(\epsilon_N)$.

On the other hand, the budget manager can return $(0, 0, +\infty)$ which causes
the prediction to be reported without spending any budget. This decision could
be based on any public information that gives high confidence to the prediction.
A good use of this case can be found in Section
\ref{experiments-skip-the-test} where \emph{timing information} is
used to skip the test.

Note that the prediction is computed from public knowledge, so releasing it has
no privacy cost. However in this case we loose any guarantee on the utility of
the reported answer, at least in the general case; based on the criteria for
skipping the test (as in the case of the user walking in the city), we could
make assumptions about the quality of the prediction which would allow
to restore the bound.

Note also that a purely predictive mechanism could be a viable alternative also
when the mechanism runs out of budget and should normally stop.
Reporting an untested prediction for free could provide some utility in this
case.

\vspace{-5pt}
\section{Predictive mechanism for location privacy}\label{sec:predictive-geo-ind}
The applicability of \priv{\dx} to location-based systems, called geo-indistinguishability
in this context, was already discussed in Section~\ref{sec:preliminaries}.
Having studied the general properties of our predictive mechanism,
we are ready to apply it for location privacy.

As already described in the preliminaries the sets of secret and
observables are sets of geographical coordinates, the metric used is
the euclidean distance and we will use $\Theta(\epsilon_\theta,l,\tilde{z})$ 
\eqref{eq:laplacetest} as test function.
We start with the description of a simple prediction function,
followed by the design of two budget managers and finally some
heuristics used to skip the test.

\vspace{-8pt}
\subsubsection*{Prediction Function.}
For the prediction function we use a simple strategy, the
$\mathtt{parrot}$ prediction, that just returns the value of the last
observable, which ultimately will be the last hard observable.
\begin{equation}
	\mathtt{parrot}((z,b)::\vc{r}) = z\label{eq:3}
\end{equation}
Despite its simplicity, this prediction gives excellent results in the case
when
the secrets are close to each other with respect to the utility
required - e.g. suppose the user queries for restaurants and he is
willing to accept reported points as far as 1 km from the secret point,
if the next positions are tens of meters apart, then the same reported
point will be a good prediction for several positions. Similarly,
the prediction is quite effective when
the user stays still for several queries, which is a typical case
of a smartphone user accessing an LBS.

More concretely, we define the \emph{step} of a trace as the average
distance between its adjacent points $\sigma (\vc x) = \mathrm{avg}_{0
\leq i < |\vc x|}\; d(x_i,x_{i+1})$ and we compare it with the
$\alpha_N(0.9)$-accuracy of the noise mechanism.
The intuition is that the parrot prediction works well on a trace $\vc
x$ if $\sigma(\vc x)$ is smaller than $\alpha_N(0.9)$ or in the presence of
clusters because once we release a hard point we can use it as a good
enough prediction for several other secret points close to it.

Furthermore the parrot prediction can be trivially implemented on any
system and it has the desirable property of being independent from the
user; taking into account past traces of the user, for instance, would
give a more effective prediction, but it would be restricted to that
particular user.


\vspace{-8pt}
\subsubsection*{Budget Managers}
When configuring a mechanism we need to take into account
3 global parameters: the global privacy, the utility and the
number of interactions, written $(\epsilon, \alpha, n)$ for 
brevity.
All three are interdependent and fixing one we obtain a relation
between the other two.
In our case we choose to be independent of the length of the traces;
to do so we introduce the \emph{privacy consumption rate}
(or just rate) which is the amount of budget spent at each
step on average: $\rho(\vc{r}) = \frac{\epsilon(\vc r)}{|\vc r|}$.
This measure represent the privacy usage of the mechanism or how
\emph{fast} we run out of budget and given this value we can easily
retrieve how many points we can cover given a certain initial budget.
As already done for $\derr$, we also introduce the average-case rate
for the mechanism as the expected value of $\rho$, given a
prior distribution $\pi\in\calp(\calx^n)$ on traces:
\[
	E[\rho] = \displaystyle 
		\smallsum{\vc{x}} \pi(\vc{x})
		\smallsum{\vc{r}} \PM(\vc{x})(\vc{r}) \; \rho(\vc{r})
\]
Given that our main concern is privacy we restrict ourselves to
$\epsilon$-\emph{bounded} budget managers, that guarantee that the
total budget consumed by the mechanism will never exceed $\epsilon$,
and divide them in two categories:

\emph{Fixed Utility:}
In the independent mechanism if we want to guarantee a certain level
of utility, we know that we need to use a certain amount of budget at
each step, a fixed rate, thus being able to cover a certain number $n$
of steps.
However in our case, if the test is successful, we may save the cost
of the noise and meet the fixed utility with a smaller rate per point; 
smaller rates translates in additional interactions possible after
$n$.  
We fix the utility and minimize the rate.

\emph{Fixed Rate:}
Alternatively, if in the independent mechanism we want to cover just
$n$ steps, thus fixing the rate, we would obtain a certain fixed
utility.
On the contrary the predictive mechanism, in the steps where the test
succeeds, spends less than the chosen rate, allowing the next steps to
spend \emph{more} than the rate.
This alternance creates a positive behavior where hard points can use
the saved budget to increase their accuracy that in turn makes
predicting more accurate and likely to succeed, leading to more
saving.
Of course the average cost for all steps meets the expected rate.
In this case we fix the rate and maximize the utility. 

In both approaches (and all strategies in between), it is never easy to
determine exactly the behavior of the mechanism, for this reason the
budget manager should always be designed to respond dynamically over
time.

\vspace{-8pt}
\subsubsection*{Configuration of the mechanism}\label{sec:configuration}
We now give an overview of the constraints that are
present on the parameters of the predictive mechanism and a guideline
to configure them to obtain the desired levels of privacy and
utility.
The only settings that the user needs to provide are $\epsilon$ and
either $\alpha$ or $\rho$.
The budget manager will define at each step the amount of budget
devoted to the test $\epsilon_{\theta}$, the noise mechanism
$\epsilon_N$ and the test threshold $l$, starting from the global
settings.

\vspace{-5pt}
\paragraph{Budget usage}
First we define the \emph{prediction rate} $PR$ as the
percentage points predicted successfully; this property will be used to
configure and to verify how effective is the predictive mechanism.
We can then introduce a first equation which relates $\epsilon_{\theta}$ and
$\epsilon_N$ to the budget consumption rate: 
$\rho = \epsilon_{\theta} + (1-PR) \epsilon_N$.
This formula is derived from the budget usage of the mechanism
(Lemma \ref{th:privacy}), with the two following approximations.
First, $\epsilon_{\theta}$ and $\epsilon_N$ in future steps are assumed
constant. In practice they will be variable because this computation is
re-done at each step with the actual remaining budget.
Second, we assume the hard steps are evenly distributed along the run.
This allows us to use PR, which is a global property of the trace,
in a local computation.

Note that $\rho$ is constant in the fixed rate case and is computed
over the current run for the fixed utility case.
We already knew that the budget available at each step had to be split
between $\Theta$ and $N$, this result confirms the intuition that the
more we manage to predict (higher PR) the less we'll need to spend for
the noise generation (on average over the run).

\vspace{-5pt}
\paragraph{Utility}
From the utility result given by Proposition~\ref{th:accuracy} we
obtain an equation that relates all the parameters of the mechanism,
$\epsilon_{\theta}$, $\epsilon_N$ and $l$.
Given that the global utility will be the worst of the two, we decide
to give both the noise and predictive components the same
utility: $\alpha_N = l + \alpha_{\theta}$.
Moreover, as discussed in the utility section, this result is a
bound valid for every possible prediction function, even one that
always fails, for this reason the bound may be too pessimistic for the
practical cases where the prediction does work.
In order to reduce the influence of the accuracy of the predictive
component we introduce a parameter $0 \leq \eta \leq 1$ that can be
set to $1$ to retrieve the strict case or can safely go as low as
$0.5$ as shown in our experiments. Finally we obtain the following 
relation between the parameters:
$\alpha = \alpha_N = \eta (l + \alpha_{\theta}) \label{eta}$.

\vspace{-5pt}
\paragraph{Noise-threshold ratio}
Now we have two equations for three parameters and to completely
configure the mechanism we introduce an additional parameter $0 \leq
\gamma \leq 1$ that is used to tune, in the predictive component, the
ratio between the threshold $l$ and the Laplacian noise added to it 
so that $\gamma = \frac{\alpha_{\theta}}{l}$.
The intuition is that $\gamma$ should not be bigger that 1, otherwise
the noise could be more important than the threshold and we might as
well use a random test. For our experiments we found good values of
$\gamma$ around $0.8$.

Note that both $\eta$ and $\gamma$ are values that should be determined
using a representative sample of the expected input, in a sort of tuning
phase, and then fixed in the mechanism. 
The same goes for the expected prediction rate that is used to
configure the budget managers, at least in the beginning this value is
necessary to allocate some resource for $\Theta$, after some
iterations it is computed from the actual run.

\setlength{\belowcaptionskip}{-12pt}
\begin{figure*}[t]
  \begin{subfigure}[b]{0.45\textwidth}

\begin{lstlisting}[mathescape, language=C, emph={mechanism,then,budget,manager,STOP},emphstyle=\textbf]
budget manager $\beta$($\rb$)
  if $\epsilon(\vc{r}) \geq \epsilon$ then STOP 
  else
    $\epsilon_{\theta} := \eta \frac{c_{\theta}}{\alpha} (1 + \frac{1}{\gamma})$
    $\epsilon_N := \frac{c_N}{\alpha}$
    $l := \frac{c_{\theta}}{\gamma  \epsilon_{\theta}}$
    return $(\epsilon_{\theta},\epsilon_N,l)$
\end{lstlisting}
\caption{Fixed Utility configured with $\epsilon$ and $\alpha$}\label{fig:fixed-u}
\end{subfigure}
~
\begin{subfigure}[b]{0.45\textwidth}

\begin{lstlisting}[mathescape, language=C, emph={mechanism,then,budget,manager,STOP},emphstyle=\textbf]
budget manager $\beta$($\rb$)
  if $\epsilon(\vc{r}) \geq \epsilon$ then STOP 
  else
    $\epsilon_N := \frac{\rho}{(1-PR) + \frac{c_{\theta}}{c_N} \eta (1 + \frac{1}{\gamma})}$
    $\epsilon_{\theta} := \epsilon_N \eta \frac{c_{\theta}}{c_N} (1+\frac{1}{\gamma})$
    $l := \frac{c_{\theta}}{\gamma \epsilon_{\theta}}$
    return $(\epsilon_{\theta},\epsilon_N,l)$
\end{lstlisting}
\caption{Fixed Rate configured with $\epsilon$, $\rho$ and $PR$}\label{fig:fixed-rate}
\end{subfigure}
\end{figure*}
\setlength{\belowcaptionskip}{0pt}

\vspace{-5pt}
\paragraph{Relation between accuracy and epsilon}
The final simplification that we apply is when we compute the accuracy
of the noisy components, for both the linear Laplacian and the polar
Laplacian we can compute their maximum value up to a certain
probability $\delta$ using their inverse cumulative probability
distributions, that we denote $\texttt{icll}$ and $\texttt{icpl}$
respectively.
Fixing $\delta$ to $0.9$, both these functions can be expressed as the
ratio of a constant and the epsilon used to scale the noise 
$\alpha_N(\delta) = \mathtt{icpl}(\epsilon_N,\delta) = \frac{c_N(\delta)}{\epsilon_N}$ and 
$\alpha_{\theta}(\delta) \;= \mathtt{icll}(\epsilon_{\theta},\delta) \;= \frac{c_{\theta}(\delta)}{\epsilon_{\theta}}$.


Note that this characterization of $\alpha_N$ is valid only for the
polar Laplacian used to achieve geo-indistinguishability. In fact, this is the only
domain specific part of the configurations presented, that can
otherwise be applied to a generic notion of \priv{\dx}.

Now that we have the equations that relate the various parameters,
from the settings given by the user we can realize the two budget
managers, shown in Figure \ref{fig:fixed-u} and
\ref{fig:fixed-rate}.

Furthermore we can compare the expected rate or accuracy of our
mechanism with those of an independent mechanism and find the
prediction rate that we need to meet to provide an improvement. We
obtain in both cases a lower bound on the prediction rate: $PR \geq
\eta \frac{c_{\theta}}{c_N} (1 + \frac{1}{\gamma})$.
This gives an idea of the feasibility of a configuration before
actually running it, for example using the parameters of our
experiments we find that it is necessary to predict at least 46\% of
points to make up for the cost of the test.

\vspace{-5pt}
\section{Case study}\label{sec:case-study}

To evaluate our mechanism, we follow our motivating example stated in the
introduction of a user performing several activities while moving around the
city throughout a day, possibly using different means of transport. During these
activities, the user performs queries to an LBS using his mobile device, while
wishing to keep his location private.

We assume that the user queries the LBS only when being still or
moving at a slow speed (less than 15 km/h); this reflect the semantic
of a geo localized query: there is usually little value in asking information
relative to one's current position if the position is changing quickly.
We perform a comparison between the independent
mechanism $\IM$ and our predictive mechanism $\PM$,
both using polar Laplace noise as the underlying noise mechanism.
The mechanisms are evaluated on
two data sets of real GPS trajectories, using both a fixed-utility
and fixed-rate budget managers and a skip strategy.

\vspace{-5pt}
\paragraph{Data sets}
The first data set we tested our mechanism against, is the well known
GeoLife \cite{DBLP:journals/debu/ZhengXM10} which collects 18.670 GPS
trajectories from 182 users in Beijing during a period of over five
years.
In this set the users take a variety of means of transport, from
walking and biking to car, train, metro, taxi and even airplane.
Regarding the trajectories length we can roughly divide them on three
equal groups, less than 5 km, between 5 and 20 km and more than 20 km.
As for duration 58\% are less than 1 hour, 26\% between 1 and 6 hours
and 16\% more than 6 hours.

The second data set is Tdrive \cite{DBLP:conf/gis/YuanZZXXSH10}, a
collections of about 9000 taxi trajectories, always in the city of
Beijing. 
As opposed to the variety of Geolife in this set we have only cars
movements and the trajectories tends to be longer in both time and
distance.
The interest of using this set, which does not exactly correspond to
our target use case of a user walking in a city, is to test the
flexibility of the mechanism.

In order to use this sets some preprocessing is needed in order to
model our use case.
GPS trajectories present the problem of having \emph{all
the movements} of the user, instead of just the points where the user
actually \emph{queried} the LBS, which is a small subset of the trajectory.
For this reason we perform a probabilistic ``sampling'' of the trajectories
that, based on the speed and type of user, produces a trace of query
points.
First, we select the part of the trace where the speed is less
than 15 km/h, and in these segments we sample points depending on the type
of user, as explained below.

Users are classified based on the frequency of their use of the LBS, from
occasional to frequent users. This is achieved by defining two intervals in
time, one brief and the other long (a \emph{jump}), that could occur between two
subsequent queries. Then each class of users is generated by sampling with a
different \emph{probability of jumping} $p$, that is the probability that the
next query will be after a long interval in time. Each value of $p$ gives rise
to a different prior distribution $\pi$ on the produced traces, hence affecting
the performance of our mechanism.

The interval that we used in our experiments are 1 and 60 minutes,
both with addition of a small Gaussian noise; frequent users will
query almost every minute while occasional users around every hour.
In our experiments we generated 11 such priors, with probability of
jumping ranging from 0 to 1 at steps of 0.1, where each trace was
sampled 10 times.

\vspace{-5pt}
\paragraph{Configuration}
In order to configure the geo-indistinguishable application, first the
user defines a radius $r^*$ where she wishes to be protected, that we
assume is 100 meters, and then the application sets $\epsilon^*$, the
global level of privacy, to be $\ln 10$. 
This means that taken two points on the radius of 100 meters their
probability of being the observables of the same secret differ at most
by $10$, and even less the more we take them closer to the secret.
We think this is a reasonable level of privacy in a dense urban
environment.
For what concerns the two budget managers, the fixed-rate was tested
with a $3.3\%$ rate, which corresponds to about 30 queries, which in a
day seems a reasonable number even for an avid user.
For the fixed-utility we set an accuracy limit 3 km, again reasonable
if we consider a walking distance and that these are worst cases.

\vspace{-5pt}
\paragraph{Skip-the-test strategy}\label{experiments-skip-the-test}
While the aim of the mechanism is to hide the user's position, the
timestamp of a point is observable, hence we can use the elapsed time
from the last reported point to estimate the distance that the user
may have traveled.
If this distance is less than the accuracy required, we can report the
predicted value without testing it, we know that the user can't be too
far from his last reported position.
The risk of this approach lies in the speed that we use to link
elapsed time and traveled distance, if the user is faster that
expected (maybe he took a metro) we would report an inaccurate point.
To be on the safe side it should be set to the maximum speed we expect
our users to travel at, however with lower values we'll be able to
skip more, it is a matter of how much we care about accuracy or how
much we know about our users.
In our experiments we assumed this speed to be 0.5 km/h.

We would expect this approach to be more convenient in a context where
accuracy is not the primary goal; indeed skipping the test will
provide the greatest advantage for the fixed-utility case, where we
just don't want to exceed a worst case limit.

Additionally we use another skip-the-test strategy to use directly with
the noise mechanism when we are in the first step and thus
there is no previous hard point for the parrot prediction to report.
This is a trivial example of skip strategy, yet it can lead to some budget
savings.

\setlength{\textfloatsep}{7pt}
\floatstyle{boxed}
\restylefloat{figure}
\begin{figure*}[t]
  \centering
  \begin{subfigure}[b]{0.92\textwidth}
    \centering
    \includegraphics[width=0.95\columnwidth]{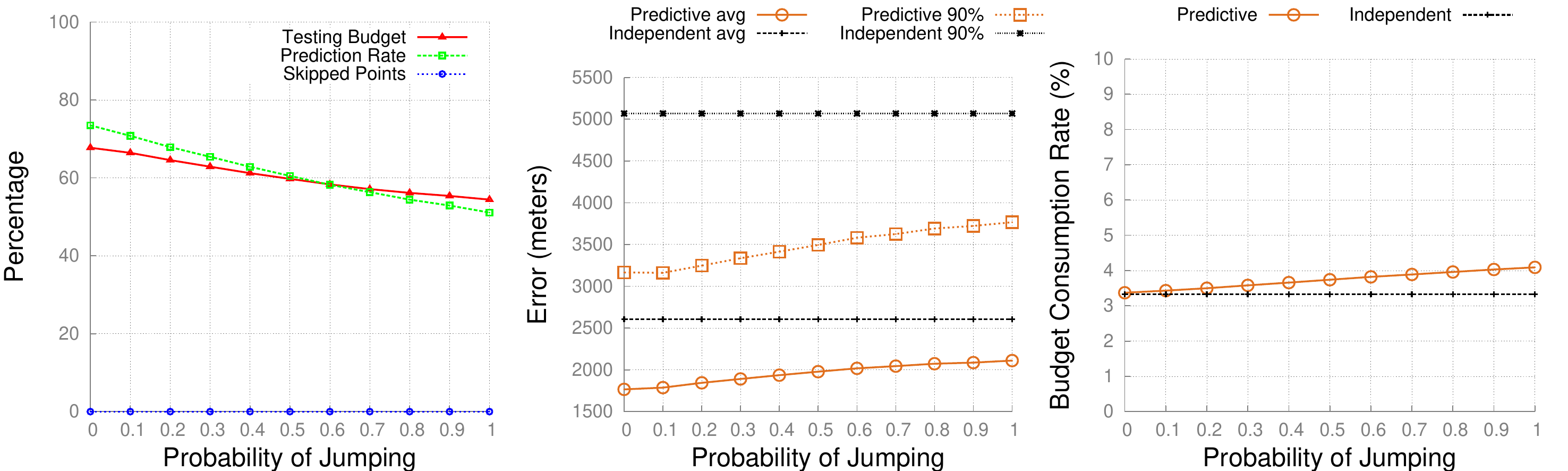}
    \caption{Fixed-Rate 3\% without skip}
  \end{subfigure}
  \\
  \begin{subfigure}[b]{0.92\textwidth}
    \centering
    \includegraphics[width=0.95\columnwidth]{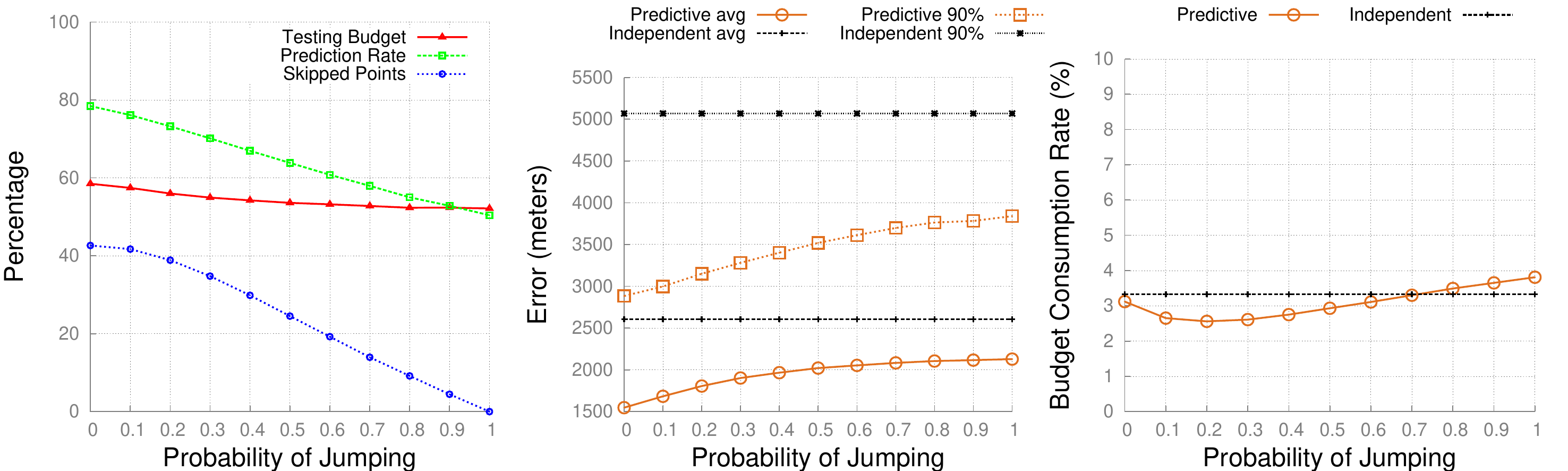}
    \caption{Fixed-Rate 3\% with skip}
  \end{subfigure}
  \caption{General statistics, Average Error and Rate for Fixed-Rate budget manager.}\label{fig:results-fixed-r}
\end{figure*}

\begin{figure*}[t]
  \centering
  \begin{subfigure}[b]{0.92\textwidth}
    \centering
    \includegraphics[width=0.95\columnwidth]{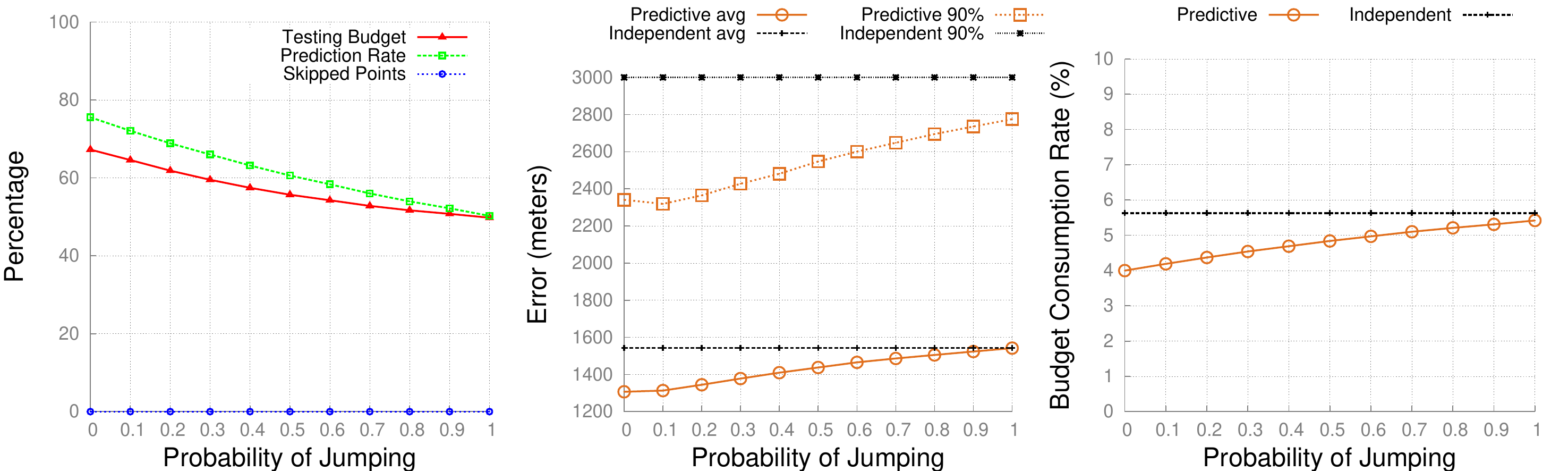}
    \caption{Fixed-Utility 3 km without skip}
  \end{subfigure}
  \\
  \begin{subfigure}[b]{0.92\textwidth}
    \centering
    \includegraphics[width=0.95\columnwidth]{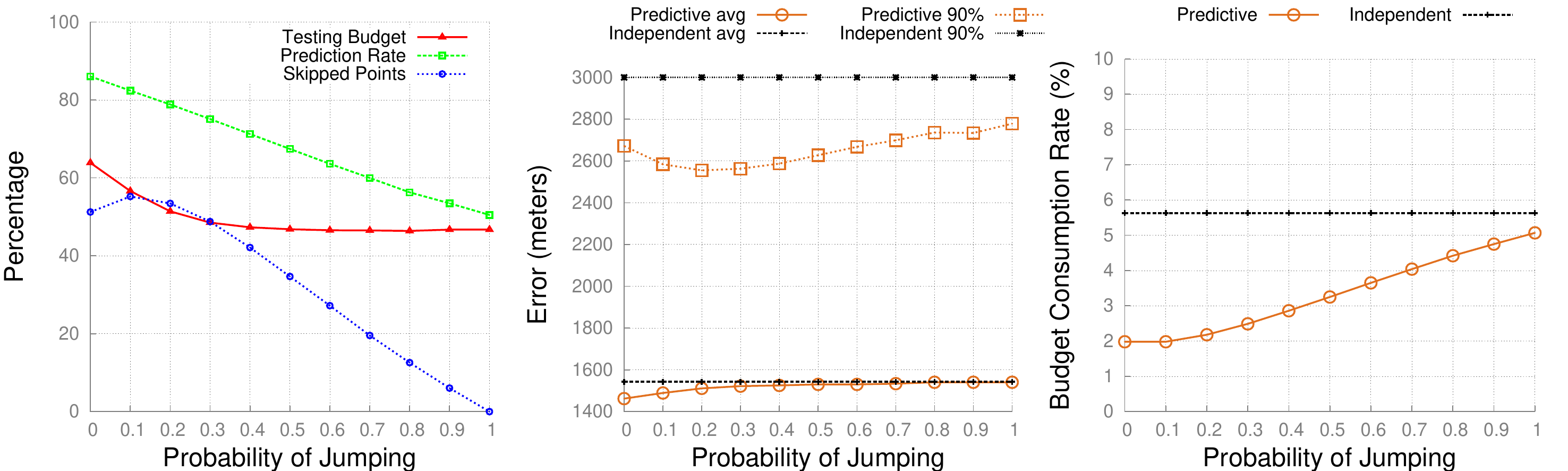}
    \caption{Fixed-Utility 3 km with skip}
  \end{subfigure}

  \caption{General statistics, Average Error and Rate for Fixed-Utility budget manager.}\label{fig:results-fixed-u}
\end{figure*}

\vspace{-5pt}
\paragraph{Results}
It should be noted that both the preprocessing and the sanitization
were performed with same configuration on both data sets.
The results of running the mechanism on the samples traces from the
Geolife data set, are reported in figures \ref{fig:results-fixed-r},
\ref{fig:results-fixed-u}, the graphs of Tdrive are omitted for reason
of space as they show a very similar behavior to Geolife (they can be
found in the appendix \ref{omitted-graphs}).
In the horizontal axis we have the probability $p$ that was used during
the sampling, to determine how often the user performs a \emph{jump}
in time: the smaller the value the more frequent the queries.
For each budget manager we plot: In the first graph, some general
statistics about the mechanism, such as the prediction rate achieved,
the amount of budget devoted to $\Theta$ and the amount of skipped points;
In the second column the average ($E[\derr]$) and 90-th percentile
($\alpha(0.9)$) of the error; In the third the average budget
consumption rate $E[\rho]$.
Furthermore we run the experiments with and without the skip the test
strategy, for the sake of comparison.

The graphs present a smooth behaviour, despite the use of real data,
because of the sampling on each trace and the averaging over all
traces of all users.
As general remarks, we can see that the prediction rate degrades as
the users become more occasional, thus less predictable, and the same
goes for the number of skipped points.
Notice that the testing budget adapts with the prediction rate which
is a sign that the budget managers reconfigure dynamically.

\emph{Fixed-rate (Fig. \ref{fig:results-fixed-r})}: fixing the
rate to $3.3\%$ to cover 30 points, we can devote the budget saved to
improve the accuracy.
In the right most graph we see that indeed the rate is very stable
even in the unpredictable cases, and very close to the rate of the
independent mechanism.
The graph in the center shows great improvements in the average error,
500 m in the worst case and 700 m in the best, and even more
remarkable is the improvement for the maximum error, 1.3km up to 1.9km.
With the skip strategy we see a small improvement for $p \leq 0.5$,
again both in average and maximum error, which correspond to a
decrease in the testing budget in the left most graph: the budget
saved skipping the test is invested in more accurate noise.

\emph{Fixed-utility (Fig. \ref{fig:results-fixed-u})}: fixing the
maximum utility (or in-accuracy) to 3 km, our mechanism manages to
save up to 1.5\% of budget rate.
If we want to compare the number of points covered, the independent
mechanism can do around 17 points while the predictive 24.
As expected the average and max errors are below the independent
mechanism corresponding values which confirms that the budget manager
is working correctly keeping the utility above a certain level.
Despite this they don't show a stable behavior like the rate in the
fixed-rate case, this is due to the fact that while we can finely
control the amount of budget that we spend, the error is less
controllable, especially the one produced by the predictive component.
With the skip strategy in this case we obtain a very noticeable
improvement in this case, with rates as low as $2\%$ in the best case
which translates to 50 points covered.
As already pointed out, in this case the skip strategy is more fruitful
because we care less about accuracy.

\vspace{-5pt}
\paragraph{Tdrive} 
This data set reports remarkably similar performance to Geolife when
the probability of jumping $p$ is less than 0.7.
In this cases the predictive mechanism is consistently a better choice
than the independent mechanism on both budget managers.
On the contrary for higher values of $p$ the independent mechanism
performs better, it is interesting to notice that the prediction rate
at $p=0.7$ starts to be lower than $46\%$, as expected from Section
\ref{sec:configuration}.
This difference between the best and worst case is more accentuated in
Tdrive precisely because the prediction function was not designed for
this scenario.
The more sporadic users are even less predictable as they are moving
at higher speeds and roaming larger areas.
Also the skip strategy, again designed for walking users, shows some
spikes in the average error, due to wrongly skipped points where
probably the taxi speeded up suddenly.

\begin{wrapfigure}[24]{R}{0.40\textwidth}
  \vspace{-20pt}
  \centering
  \includegraphics[width=0.40\textwidth]{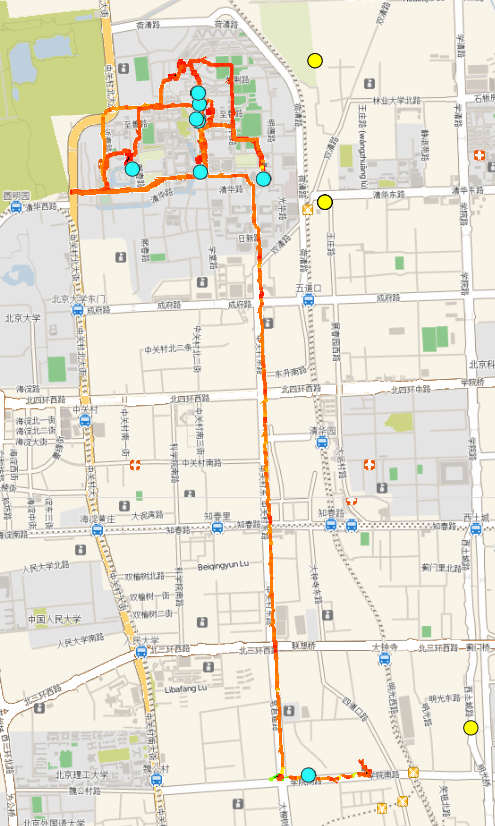}
  \caption{Original trace (red), sampled trace (light blue) and reported trace
  (yellow).}
  \label{fig:trace}
\end{wrapfigure}

Figure \ref{fig:trace} displays one of Geolife trajectories sanitized
with fixed utility.
The original trace, in red, starts south with low speed, moves
north on a high speed road and then turns around Tsinghua University for
some time, again at low speed, for a total of 18 km traveled in 10
hours.
The sampled trace was obtained with a probability 0.5 of jumping and
is plotted in light blue: as expected, 9 of the points are north, one
south and the middle part was skipped.
Finally in yellow we have the reported trace with 3 locations, which
were used once for the point at the bottom, 7 times for the one in the
middle and twice for point in the top.

\floatstyle{plain}
\restylefloat{figure}

\vspace{-5pt}
\section{Conclusions and Future Work}
\label{sec:conclusions}
\vspace{-5pt}
\subsubsection*{Future Work.}
As the experiments show the more efficient use of budget allows us to
cover a day of usage, which was the goal we were aiming for in order
to attack realistic applications.
The intuition is that even if there is correlation between the traces of
the same user on several days (for example we go to work and home
every day) still it is not enough to accurately locate the user at a
precise moment in time (we might go to work later, or follow a
different road).
It is not clear though if one day is enough time to break the
correlation and possibly reset the budget, we leave to future work to
investigate in which cases it is indeed possible to reset the system
and when on the contrary the epsilon keeps increasing.

One other possibility to prolong even further the use of the system is
to improve the prediction.
The experimental part of this paper was carried on with a prediction
simple enough to be effective yet not distracting with engineering
details.
An extension we plan to develop consist in using the mobility
traces of a user, or of a group of users, to designate locations where
the next position is likely to be.
In \cite{DBLP:journals/tdp/GambsKC11} the authors already developed
inference attacks on the reported locations of users to discover
points of interests and future locations, among other things; the idea
is to use these attacks as a prediction.
If we consider the use case of a mobile phone, the mechanism itself
could collect the reported traces and train itself to
predict more accurately.

We are also developing a \emph{linearizing} prediction, that
determines the direction using a linear regression method which
additionally allows to detect \emph{turns}, sharp changes in
direction, thanks to the error reported.
This kind of prediction targets cases where the system needs to
frequently report its position with good accuracy, such as a
navigation system, we think that a small amount of privacy could still
be desirable, for example to hide the exact position along a road.
Of course this prediction only works in cases where the secret
trajectory is very linear, restricting its usage to cases such as
trains, airplanes or possibly boats as means of transport.
One possible improvement could be the use of a non-linear regression
technique but it still has to be explored.

Alternatively we are considering the use of public geographic
information to improve the prediction, which could simply translate to
using already developed map-matching algorithms: typically in
navigation systems an approximate location needs to be matched to an
existing map, for example to place the user on a road.
Map matching would make trivial predicting the direction of the user
moving on a road for example, while in crossroads could be dealt with
with the help of the mobility traces already discussed before: if on
the left the is just countryside and on the right a mall, the user is
more likely to turn right.
Ultimately if more than one prediction function prove effective, we are
interested in the possibility to merge them, for instance using multiplicative
weights or related technique (e.g. Kalman filters): each prediction is
assigned a weight, at each turn the prediction with the highest weight
is interrogated, if we are in easy case its weight is raised otherwise
is reduced, in the hope that when changing scenario the weights would
adjust and the right prediction would be picked.

\vspace{-5pt}
\subsubsection*{Related work.}
On the predictive mechanism side, our mechanism was mainly inspired
by the median mechanism \cite{Roth:10:STOC}, a work on differential
privacy for databases based on the idea of exploiting the
correlation on the queries to improve the budget usage. 
The mechanism uses a concept similar to our \emph{prediction} to
determine the answer to the next query using only past answers.
An analogous work is the multiplicative weights mechanism
\cite{DBLP:conf/focs/HardtR10}, again in the context of statistical
databases. 
The mechanism keeps a parallel version of the database which is used
to predict the next answer and in case of failure it is updated with a
multiplicative weights technique.

A key difference from our context is that in the above works, several queries
are performed against the \emph{same database}. In our setting, however, the
secret (the position of the user) is always changing, which requires to exploit
correlations in the data. This scenario
is explored also in \cite{DBLP:conf/stoc/DworkNPR10} were the authors consider
the case of an evolving secret and develop a differentially private counter.

Concerning location privacy, there are excellent works and surveys
\cite{Terrovitis:11:SIGKDD,Krumm:09:PUC,Shin:12:WC} that present the
threats, methods, and guarantees.
Like already discussed in the introduction the main trends in the
field are those based on the expectation of distance error
\cite{Shokri:11:SP,Shokri:12:CCS,Hoh:05:SecureComm,Dewri:12:TMC} and
on the notion of $k$-anonymity
\cite{Kido:05:ICDE,Shankar:09:UbiComp,Bamba:08:WWW,Duckham:05:Pervasive,Xue:09:LoCa},
both dependents on the adversary's side information, as are some other
works \cite{Cheng:06:PET} and \cite{Ardagna:07:DAS}.

Notions that abstract from the attacker's knowledge based on
differential privacy can be found in \cite{Machanavajjhala:08:ICDE}
and \cite{Ho:11:GIS} although only for \emph{aggregate} information.

The notion we based our work on, geo-indistinguishability
\cite{Andres:13:CCS}, other than abstracting from the attacker's prior
knowledge, and therefore being suitable for scenarios where the prior
is unknown, or the same mechanism must be used for multiple users, can
be used for single users.
In addition, being the definition an instantiation of the more general
notion of \priv{\dx} \cite{Chatzikokolakis:13:PETS} we were able to
generalize our mechanism as well, being the prediction the only domain
specific component.

\vspace{-5pt}
\subsubsection*{Conclusions.}
We designed a general framework for private predictive
\privadj{\dx} mechanisms able to manage the privacy budget more
efficiently than the standard approach, in the cases where there is a
considerable correlation on the data.
The mechanism is modular and clearly separates the privacy protecting
components from the predictive components, allowing ease of analysis
and flexibility.
We provide general configuration guidelines usable for any notion of
\priv{\dx} and a detailed instantiation for geo indistinguishability.
We tested the geo private mechanism obtained with two large sets of GPS
trajectories and confirmed the goals set in the design phase.
Experimental results show that the correlation naturally present in a
user data is enough for our mechanism to outperform the independent
mechanism in the majority of prior tested.

\vspace{-5pt}
\bibliographystyle{splncs}
\bibliography{short}

\newpage
\appendix

\section{Graphs of experiments on Tdrive data set }\label{omitted-graphs}

\floatstyle{boxed}
\restylefloat{figure}
\begin{figure*}[h!]
  \centering
  \begin{subfigure}[b]{0.98\textwidth}
    \centering
    \includegraphics[width=\columnwidth]{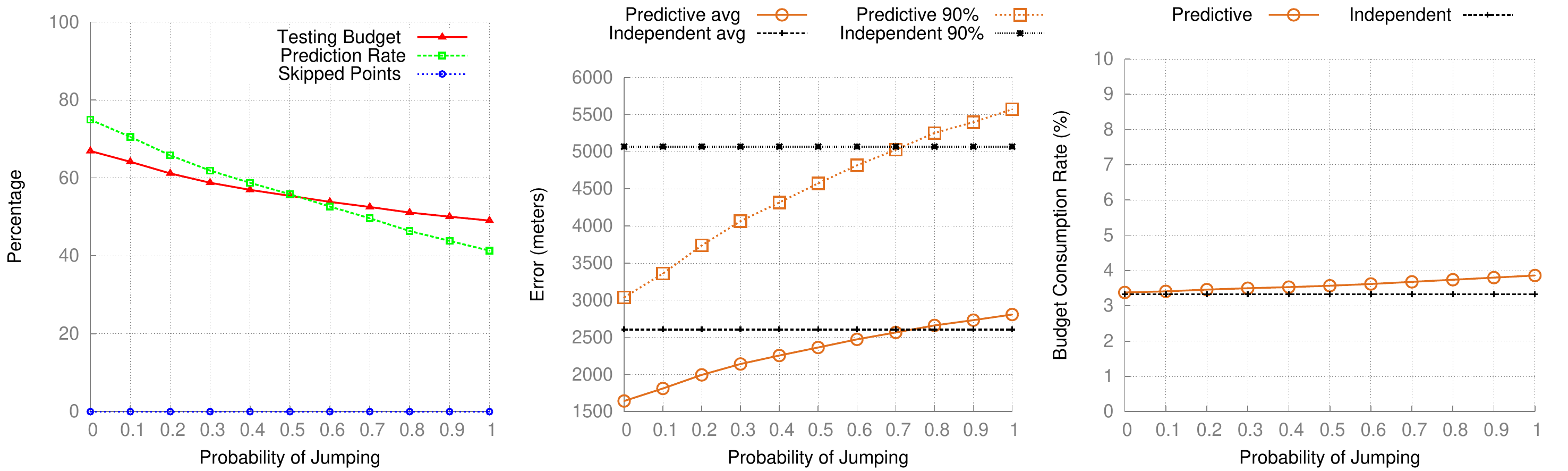}
    \caption{Fixed-Rate 3\% without skip}
  \end{subfigure}
  \\
  \begin{subfigure}[b]{0.98\textwidth}
    \centering
    \includegraphics[width=\columnwidth]{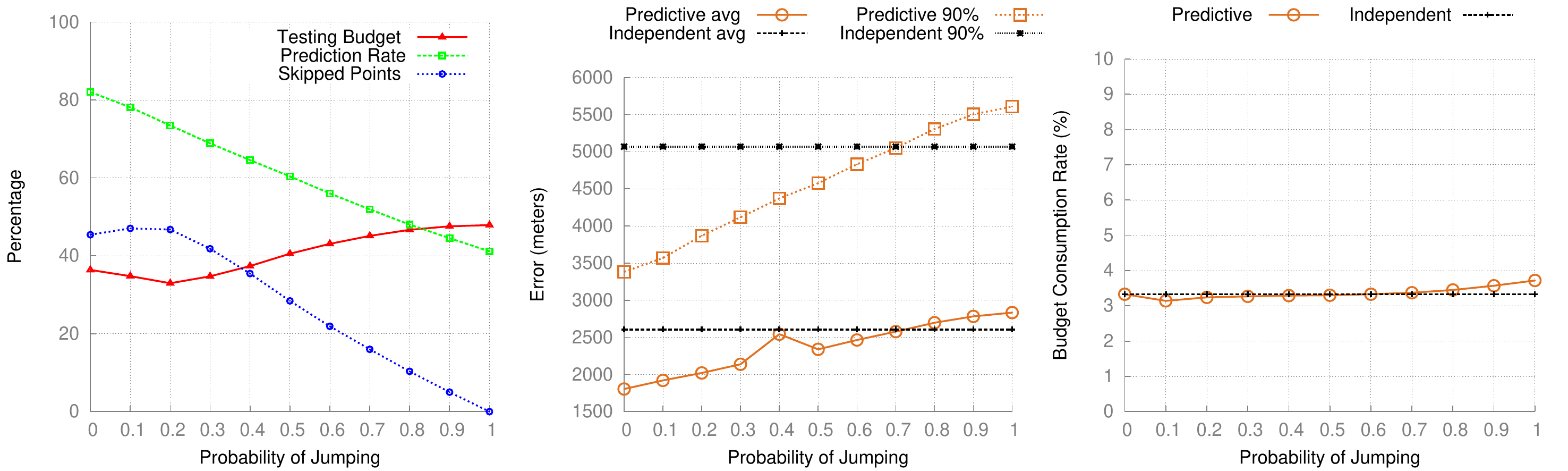}
    \caption{Fixed-Rate 3\% with skip}
  \end{subfigure}
\\
  \begin{subfigure}[b]{0.98\textwidth}
    \centering
    \includegraphics[width=\columnwidth]{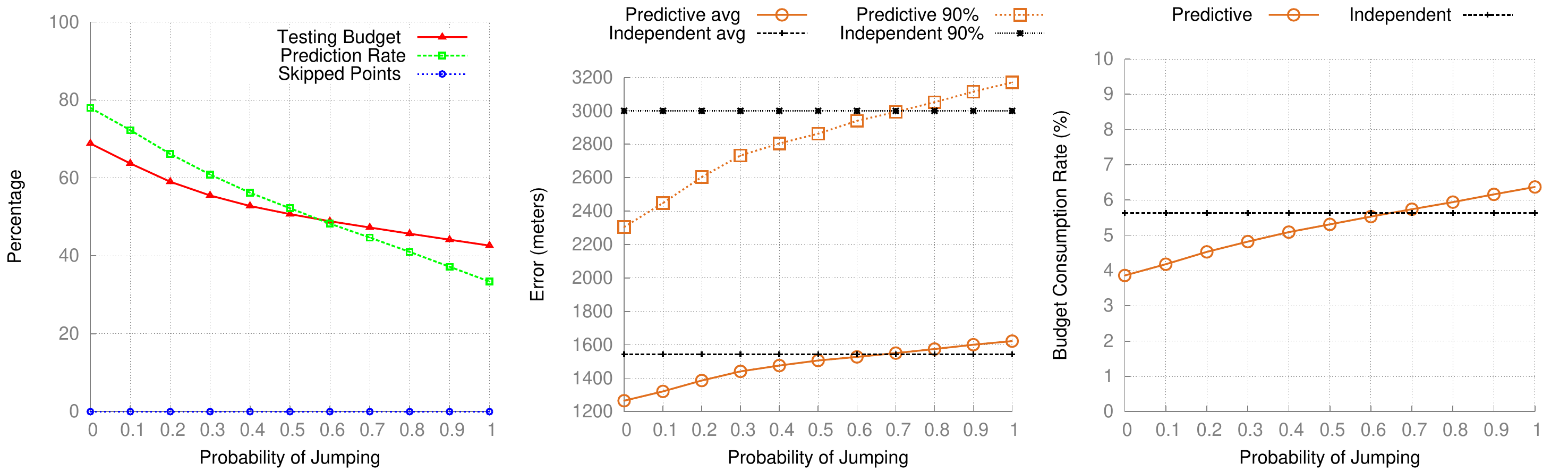}
    \caption{Fixed-Utility 3 km without skip}
  \end{subfigure}
  \\
  \begin{subfigure}[b]{0.98\textwidth}
    \centering
    \includegraphics[width=\columnwidth]{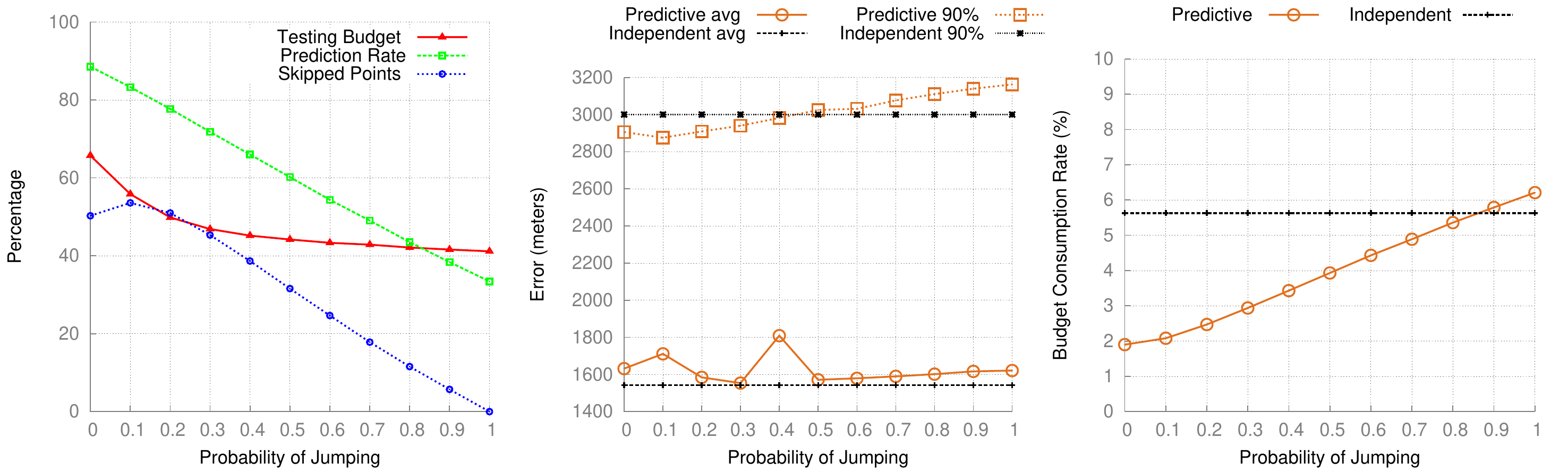}
    \caption{Fixed-Utility 3 km with skip}
  \end{subfigure}

  \caption{Tdrive: General statistics, Average Error and Rate.}\label{fig:tdrive-results}
\end{figure*}

\newtheorem{innercustomfact}{Fact}
\newenvironment{custom-fact}[1]
  {\renewcommand\theinnercustomfact{#1}\innercustomfact}
  {\endinnercustomfact}
\newtheorem{innercustomlemma}{Lemma}
\newenvironment{custom-lemma}[1]
  {\renewcommand\theinnercustomlemma{#1}\innercustomlemma}
  {\endinnercustomlemma}
\newtheorem{innercustomprop}{Proposition}
\newenvironment{custom-proposition}[1]
  {\renewcommand\theinnercustomprop{#1}\innercustomprop}
  {\endinnercustomprop}

\section{Proofs}
We provide here the proofs of all results in the paper.

\begin{custom-fact}{1}[Privacy of Test function]
  The test function $\Theta$, equipped with a laplacian noise
  generation function $Lap$ scaled by $\epsilon_{\theta}$, is
  $\epsilon_{\theta}$ \privadj{\dx}.
  \[\Theta (\epsilon_{\theta},l,\tilde{z})(x) = \left\{ 
    \begin{array}{ll}
      0 & \text{if } d(x,\tilde z) \leq l + Lap(\epsilon_{\theta})\\
      1 & \text{ow.}
    \end{array}
  \right.\]
\end{custom-fact}

\begin{proof}
We use the fact that Laplacian noise scaled by $\epsilon_{\theta}$ is
$\epsilon_{\theta}$-d.p.
We assume that $Lap$ is scaled with $\epsilon_{\theta}$ so we omit it
in the following.

\[\begin{array}{ll}
  P[d(x, \tilde{z}) \leq l + Lap(0)] = &\text{translating the noise} \\
  P[Lap(d(x,\tilde{z}) - l) \leq 0] = & l \text{ and } \tilde{z} \text{ are constants} \\
  P[Lap(t) \leq 0] \leq & \text{assumption on Lap} \\ 
  e^{\epsilon_{\theta} \cdot d(t,t')} P[Lap(t') \leq 0] \leq & \text{using } \ref{triangular-inequality}\\
  e^{\epsilon_{\theta} \cdot d(x,x')} P[Lap(t') \leq 0] = & \text{translating back} \\
  e^{\epsilon_{\theta} d(x,x')} P[d(x', \tilde{z}) \leq l + Lap(0)]
\end{array}\]

\begin{equation}\label{triangular-inequality}
  \begin{array}{llll}
    d(t,t') = & |d(x,\tilde{z}) - l - d(x',\tilde{z}) +l| \leq\\[0.5em]
    &|d(x,x') + d(x',\tilde{z}) - d(x',\tilde{z})| &= d(x,x')
\end{array}
\end{equation}
\qed
\end{proof}

\begin{custom-lemma}{1}
  A predictive mechanism $\PM$ that uses a family of \privadj{\dx}
  noise mechanisms, a family of \privadj{\dx} test functions and a
  budget manager $\beta$, satisfies
  \begin{equation}
    \forall \vc{r}.\; \forall \vc{x},\vc{x'}.\quad \PM(\vc{x})(\vc{r}) \leq 
    e^{\epsilon_{\beta}(\vc{r}) \cdot \dmax(\vc{x}, \vc{x'})} \PM(\vc{x'})(\vc{r})
  \end{equation}
\end{custom-lemma}

\begin{proof}
We want to show that:
\begin{equation}\label{eq:goal}
  \forall \vc{x},\vc{x}'.\quad P[ \vc{r} | \vc{x}] \leq e^{\epsilon(\vc{r}) \cdot d(\vc{x}, \vc{x}')} P[ \vc{r} | \vc{x}']
\end{equation}

In the following, we use the subscript $i$ to indicate both a tuple from
$0$ to $i$ such as $\vc{x}_i$ or the $i$-th element, such as $x_i$.
Decomposing in \emph{dependent} steps using the chain rule we obtain:
\begin{equation}\label{eq:dependent_steps}
  P[ \vc{r}_i | \vc{x}_i] = P[(z_i,b_i) | \vc{x}_i,\vc{r}_{i-1}] \cdot P[\vc{r}_{i-1} | \vc{x}_{i-1}]
\end{equation}

Analyzing the single step we have a binary choice between the easy
case, which is deterministic, and the hard case, which is
probabilistic. We introduce the random variable $B_i$ to denote the
outcome of the test at step $i$.
\begin{equation}\label{eq:binary_case}
  \begin{array}{llll}
    P[(z_i,b_i) | x_i, \vc{r}_{i-1}] = \\[0.5em]
    P[B_i=1 | x_i,\vc{r}_{i-1}] \cdot P[\Omega(\vc{r}_{i-1}) = z_i | \cancel{x_i},\vc{r}_{i-1}] + \\
    P[B_i=0 | x_i,\vc{r}_{i-1}] \cdot P[N(x_i) = z_i | x_i,\cancel{\vc{r}_{i-1}}] = \\[0.5em]
    P[B_i=1 | x_i,\vc{r}_{i-1}] \cdot 1 + \\
    P[B_i=0 | x_i,\vc{r}_{i-1}] \cdot P[N(x_i) = z_i | x_i] \\
  \end{array}
\end{equation}

The composition of such steps forms a binary tree with all the
possible runs that the test can produce; to treat this, we split the
tree in traces $\bar b$ as they are disjoint events.

\[\begin{array}{l}
  P[(\vc{z}, \vc{b}) | \vc{x}] = P[(\vc{z},\vc{b}) | \vc{x},\vc{b}] \cdot P[\vc{b} | \vc{x}] 
\end{array}\]

Now that we know the trace, we reorganize the indexes of its steps in
two groups, the easy $I_E=\{i \;|\; B_i=1\}$ and hard steps $I_H=\{i
\;|\; B_i=0\}$. 
After having applied assumptions \ref{privacy-assumption-test},
\ref{privacy-assumption-noise} we can regroup the exponents and
obtain a form close to \ref{eq:goal}.
Here follows the complete proof:

$\forall n, \forall \vc{x}, \vc{x}'.\quad P[\vc{r}_n | \vc{x}_n]$ = \\[0.5em]
\begin{tabular}{>{\footnotesize \centering}p{5.4em} >{$}l<{$}}
(chain rule)
& = P[\vc{r}_n | \vc{r}_{n-1},\vc{x}_n] \cdot P[\vc{r}_{n-1} | \vc{x}_n] \\[1em]
(independence from $x_n$)
& = P[\vc{r}_n | \vc{r}_{n-1},\vc{x}_n] \cdot P[\vc{r}_{n-1} | \vc{x}_{n-1}] \\[1em]
(iterating)
& = \displaystyle \prod^n_{i=1} P[\vc{r}_i | \vc{r}_{i-1}, \vc{x}_i] \\[1em]
(chain rule)
& = \displaystyle \prod^n_{i=1} P[z_i | \vc{z}_{i-1},\vc{b}_{i},\vc{x}_i] \cdot P[b_i | \vc{r}_{i-1}, \vc{x}_i] \\[1em]
(partitioning indexes)
& = \displaystyle \prod_{i \in I_H(r)} P[z_{i} | \vc{z}_{i-1},\vc{b}_{i},\vc{x}_{i}] \cdot P[b_{i} | \vc{r}_{i-1},\vc{x}_{i}] \\
& \quad \displaystyle \prod_{i \in I_E(r)} P[z_{i} | \vc{z}_{i-1},\vc{b}_{i},\vc{x}_{i}] \cdot P[b_{i} | \vc{r}_{i-1},\vc{x}_{i}]  \\[1em]
(independences)
& = \displaystyle \prod_{i \in I_H(r)}  P[z_{i} | x_{i}] \cdot P[B_{i}=0 | \vc{r}_{i-1}, \vc{x}_{i}] \\
& \quad \displaystyle \prod_{i \in I_E(r)}  1               \cdot P[B_{i}=1 | \vc{r}_{i-1}, \vc{x}_{i}]                    \\[1em]
(assumptions \ref{privacy-assumption-test}, \ref{privacy-assumption-noise})
& \leq \displaystyle \prod_{i \in I_H(r)} e^{\beta_N(\vc{r}_i) d(x,x')} \cdot P[z_{i} | x'_{i}] \cdot \\
& \quad e^{\beta_{\theta}(\vc{r}_i) d(x,x')} \cdot P[B_{i}=0 | \vc{r}_{i-1},\vc{x}'_{i}]                   \\
& \quad \displaystyle \prod_{i \in I_E(r)} e^{\beta_{\theta}(\vc{r}_i) d(x,x')} \cdot P[B_{i}=1 | \vc{r}_{i-1},\vc{x}'_{i}] \\[1em]
(grouping exponents)
& \leq e^{\epsilon(r)} \displaystyle \prod_{i \in I_H} P[z_{i} | x'_{i}] \cdot P[B_{i}=0 | \vc{r}_{i-1},\vc{x}'_{i}] \\
& \quad \qquad \displaystyle \prod_{i \in I_E} P[B_{i}=1 | \vc{r}_{i-1},\vc{x}'_{i}]                               \\[1em]
& = e^{\epsilon(\vc{r})} \cdot P[\vc{r}_n | \vc{x}'_n]
\end{tabular}

With a global exponent for the run:
\[ \epsilon(\vc{r}) = \left(\displaystyle \sum_{i \in I_H(\vc{r})} \beta_N(\vc{r}_i) + \sum_{i \in I(\vc{r})} \beta_{\theta}(\vc{r}_i) \right) \cdot \dmax(\vc{x}_n,\vc{x}'_n) \]
\qed
\end{proof}

\begin{custom-proposition}{1}[accuracy]
  Let $\vc{r}$ be a run, $\beta$ a budget manager, let
  $(\epsilon_{\theta}, \epsilon_N, l) = \beta(\vc{r})$ and let
  $\alpha_N(\delta)$, $\alpha_{\theta}(\delta)$ be the accuracy of
  $N(\epsilon_N)$, $\Theta(\epsilon_{\theta},l,\tilde{z})$ respectively.
  Then the accuracy of $\Step(\rb)$ is 
  \[ \alpha(\delta) = \max(\alpha_N(\delta), l + \alpha_{\theta}(\delta)) \]
\end{custom-proposition}

\begin{proof}
Assumptions: the noise mechanism $N$ is
$\alpha_N(\delta)$-accurate and the laplacian noise $Lap$ is
$\alpha_{\theta}(\delta)$-accurate.
The output depends on the outcome of the test function, and the
possible cases are:
\begin{itemize}
\item $A \equiv d(x,\tilde z) \leq l - |Lap(\epsilon_{\theta})|$ \\
  returns the prediction, and we know its accuracy is within $l$
\item $C \equiv l \leq d(x,\tilde z) \leq l + |Lap(\epsilon_{\theta})|$ \\
  despite it is not precise enough the prediction is returned
\item $B \equiv l - |Lap(\epsilon_{\theta})| \leq d(x,\tilde z) \leq l$ \\
  despite the prediction was precise enough, a hard point is returned,
which is $\alpha_N(\delta)$-accurate
\item $D \equiv d(x,\tilde o) \geq l + |Lap(\epsilon_{\theta})|$ \\
  returns a hard point, which is $\alpha_N(\delta)$-accurate
\end{itemize}
In the following we denote the predicate with its letter, e.g. $A$,
and the probability of it being true with $P_A$. 
In addition we denote with $H$ and $E$ the event of being in a hard or
easy case.

We want to prove that for all $\delta$, for each step $i$
\begin{equation}
  P [d(z_i,x_i) \leq \alpha(\delta)] \geq \delta \label{eq:utility}
\end{equation}

For the hard cases, we use the assumption that $N$ is
$\alpha_N(\delta)$-accurate:
\begin{equation}
  \begin{array}{ll}
    P[d(z_i,x_i) \leq \alpha_N(\delta) \;|\; H] \cdot P_H = \\[0.5em]
    P[d(z_i,x_i) \leq \alpha_N(\delta) \;|\; B] \cdot P_B + \\
    P[d(z_i,x_i) \leq \alpha_N(\delta) \;|\; D] \cdot P_D \geq \\[0.5em]
    (P_B + P_D) \delta
  \end{array}\label{eq:hard}
\end{equation}

For the easy cases, we use the assumption that $Lap$ is
$\alpha_{\theta}(\delta)$-accurate and we define the shifted accuracy
$\alpha'_{\theta}(\delta) = l + \alpha_{\theta}(\delta)$.

\begin{equation}
  \begin{array}{ll}
    P[d(z_i,x_i) \leq \alpha'_{\theta}(\delta) | E] \cdot P_E = \\[0.5em]
    P[d(z_i,x_i) \leq \alpha'_{\theta}(\delta) | A] \cdot P_A + \\
    P[d(z_i,x_i) \leq \alpha'_{\theta}(\delta) | C] \cdot P_C \geq \\[0.5em]
    1 \cdot P_A + \delta \cdot P_C
  \end{array}\label{eq:easy}
\end{equation}

We now join the two cases, choosing $\alpha(\delta) =
\max(\alpha_N(\delta),\alpha'_{\theta}(\delta))$:
\[\begin{array}{lll}
  P[d(z_i,x_i) \leq \alpha(\delta)] \geq \\[0.5em]
  P[d(z_i,x_i) \leq \alpha_N(\delta) | H] \cdot P_H + \\
  P[d(z_i,x_i) \leq \alpha'_{\theta}(\delta)|E] \cdot P_E \geq && 
  \text{using } \ref{eq:hard}, \ref{eq:easy}\\[0.5em]
  (P_B + P_D) \delta + P_A + P_C \delta = \\[0.5em]
  P_A + (P_B + P_C + P_D)\delta = \\[0.5em]
  (1-\delta) P_A + \delta \geq
  \delta
\end{array}\]
\qed
\end{proof}

\end{document}

%% file: macros.tex
\usepackage{times}
\usepackage{enumerate}
\usepackage{marginnote}
\usepackage{color}
\usepackage{url}
\usepackage{algorithm}
\usepackage{algpseudocode}
\usepackage{setspace}

\newcommand{\remove}[1]{{}}  

\newcommand{\calx}{\mathcal{X}}

\newcommand{\calz}{\mathcal{Z}}

\newcommand{\calb}{\mathcal{B}}

\newcommand{\calp}{\mathcal{P}}

\newcommand{\calr}{\mathcal{R}}

\newcommand{\PM}{\mathrm{PM}}
\newcommand{\IM}{\mathrm{IM}}
\newcommand{\Step}{\mathrm{Step}}
\newcommand{\emptytrace}{[\,]}

\newcommand{\dx}{{d_{\scriptscriptstyle\calx}}}

\newcommand{\euclid}{{d_2}}

\newcommand{\dmax}{{d_\infty}}
\newcommand{\derr}{{d_{\mathrm{err}}}}

\newcommand{\hamming}{{d_h}}

\newcommand{\dprob}{{d_\calp}}

\newcommand{\xb}{\mathbf{x}}
\newcommand{\zb}{\mathbf{z}}
\newcommand{\rb}{\mathbf{r}}

\newcommand{\reals}{\mathbb{R}}

\newcommand{\priv}[1]{$#1$-privacy}
\newcommand{\privadj}[1]{$#1$-private}

\newcommand{\edpold}[1][\epsilon]{$#1$-dif\-fer\-en\-tial privacy}

\newcommand{\geoind}{geo-indistingui\-sha\-bility}

\newcommand{\smallsum}[1]{\textstyle{\sum_{#1}\:}}

\newcommand{\smallfrac}[2]{\textstyle{\frac{#1}{#2}}}

\newcommand{\vc}[1]{ \ifmmode \mathbf{#1} \else \textbf{#1}\fi } 
\newcommand{\set}[1]{ {\cal #1} } 
\newcommand{\predz}{\tilde z}

%% file: style.tex
\usepackage[utf8x]{inputenc}   
\usepackage[dvipsnames,pdftex]{xcolor}
\usepackage[linktoc=page,colorlinks=true,urlcolor=blue]{hyperref}   
\usepackage{listings}          
\usepackage{array}             
\usepackage{enumitem}          

\usepackage[pdftex]{graphicx}  
\usepackage{wrapfig}           
\usepackage{float}
\usepackage{subcaption}   

\usepackage{amsfonts,amssymb} 
\usepackage{thmtools}          
\usepackage{thm-restate}      
\usepackage{cancel}     
\usepackage{extarrows}  
\usepackage{stmaryrd}   
\usepackage{dsfont}     
